\documentclass[conference]{IEEEtran}
\usepackage{float}
\IEEEoverridecommandlockouts
\usepackage{amsmath,amssymb,amsfonts}
\usepackage{algorithmic}
\usepackage[utf8]{inputenc}

\usepackage{graphicx}
\usepackage{textcomp}
\usepackage[utf8]{inputenc}
\usepackage[T1]{fontenc}
\usepackage[compatibility=false]{caption}
\usepackage{url}
\usepackage{hyperref}
\def\BibTeX{{\rm B\kern-.05em{\sc i\kern-.025em b}\kern-.08em
    T\kern-.1667em\lower.7ex\hbox{E}\kern-.125emX}}
    
\begin{document}

\title{Privacy-Preserving and Incentive-Driven Relay-Based Framework for Cross-Domain Blockchain Interoperability}

\author{
  \IEEEauthorblockN{
    Saeed Moradi\IEEEauthorrefmark{1},
    Koosha Esmaeilzadeh Khorasani\IEEEauthorrefmark{1},
    Sara Rouhani\IEEEauthorrefmark{1}\IEEEauthorrefmark{2}
  }
  \IEEEauthorblockA{\IEEEauthorrefmark{1}Department of Computer Science, University of Manitoba, Winnipeg, Canada \\
  Email: \{moradis1@myumanitoba.ca, esmaeilk@myumanitoba.ca, sara.rouhani@umanitoba.ca\}}
  \IEEEauthorblockA{\IEEEauthorrefmark{2}École de technologie supérieure (ÉTS), Montréal, Canada \\
  Email: sara.rouhani@etsmtl.ca}
}

\maketitle

\begin{abstract}
Interoperability is essential for transforming
blockchains from isolated networks into collaborative ecosystems,
unlocking their full potential. While significant progress
has been made in public blockchain interoperability, bridging
permissioned and permissionless blockchains poses unique
challenges due to differences in access control, architectures,
and security requirements. This paper introduces a blockchain-agnostic
framework to enable interoperability between permissioned
and permissionless networks. Leveraging cryptographic
techniques, the framework ensures secure data exchanges. Its
lightweight architectural design simplifies implementation and
maintenance, while the integration of Clover and Dandelion++
protocols enhances transaction anonymity. Performance evaluations
demonstrate the framework’s effectiveness in achieving
secure and efficient interoperability by measuring the resource
usage, the forwarding time, the throughput, and the availability
of the system across heterogeneous blockchain ecosystems.
\end{abstract}

\section{Introduction}
Blockchain technology is redefining the landscape of secure and decentralized systems, with applications spanning finance, healthcare, and supply chain management. In simple terms, a blockchain is defined as a distributed ledger structured as a linked list of blocks, each containing an ordered set of transactions. Typical implementations use cryptographic hashes to secure the linkage between a block and its predecessor \cite{xu2019architecture}. Permissioned blockchains have emerged as the backbone of enterprise solutions due to controlled access, enforceable governance policies, and enhanced privacy mechanisms tailored to regulatory and operational requirements \cite{polge2021permissioned}. Despite their advantages, these systems often operate in isolation, limiting their ability to interact with other blockchain platforms, particularly permissionless networks \cite{prusty2018blockchain, ren2023interoperability}. This lack of interoperability \cite{wang2023exploring} poses significant challenges, preventing the full realization of blockchains' potential in multi-network environments.

The challenge of achieving interoperability between heterogeneous blockchain systems has drawn considerable attention in recent years. Proposed solutions include relay chains \cite{frauenthaler_eth_2020}, hash-locking mechanisms \cite{barbara_mphtlc_2023}, and decentralized notary schemes \cite{sun_decentralized_2022}. Existing approaches usually address the concern of connecting two permissionless blockchains, leaving permissioned blockchains out of their environment. Relay chains such as Polkadot, for example, are optimized for permissionless systems but lack the privacy-preserving capabilities required for sensitive enterprise applications \cite{belchior2021survey}. Similarly, hash-locking mechanisms introduce operational complexity and high computational overhead, limiting their broader applicability in
practical settings \cite{chan2021brief}.

Moreover, privacy concerns remain largely unaddressed in existing interoperability frameworks. Permissioned blockchains typically handle sensitive data that necessitates stringent confidentiality and anonymity measures during cross-chain interactions. Without robust privacy-preserving mechanisms, enterprises face significant risks, including data leakage and compliance violations. This underscores the pressing need for a solution that combines data encryption, anonymity-preserving protocols, and seamless cross-network communication to meet the diverse requirements of heterogeneous blockchain ecosystems.

This work makes the following key contributions to the field of blockchain interoperability:
\begin {itemize}
\item \textbf{A Blockchain-agnostic Interoperability Framework}: We propose an architecture that bridges the gap between permissioned and permissionless blockchains, emphasizing privacy and security in heterogeneous environments.
\item \textbf{Privacy and Security Mechanisms}: The system leverages cryptographic methods combined with advanced routing protocols to protect user identities, obfuscate the origins of requests, and ensure data confidentiality during cross-network transactions.
\item \textbf{Security Analysis}: We conduct a formal security analysis of our relay-based architecture, demonstrating how probabilistic source-obfuscation mechanisms mitigate deanonymization attacks. By adapting Clover and Dandelion++ for cross-chain interoperability, we analyze the trade-off between anonymity and potential data exposure, providing insights into optimizing privacy-preserving transaction forwarding.
\item \textbf{Performance Evaluation}: We evaluate the framework by measuring request-response times and isolating blockchain-related processing delays, providing a focused analysis of the system’s communication efficiency and scalability.
\end {itemize}

The rest of the paper is structured as follows: Section 2 provides background on blockchain Interoperability as well as Hyperledger Fabric and Substrate, the two frameworks that we used to conduct our study. Section 3 reviews related work, identifying gaps addressed by this study. Section 4 outlines the proposed system architecture, including security and privacy mechanisms, routing protocols, and threat model. Section 5 presents evaluation along with the experimental results, highlighting the system’s performance. Finally, Section 6 concludes the study and discusses future directions.

\section {Background}

This section provides an overview of two blockchain platforms
used in this study: Hyperledger Fabric (HF) and Substrate.
Following that, we define blockchain interoperability.

\subsection{Hyperledger Fabric (HF)}  

HF \cite{androulaki_hyperledger_2018} is a permissioned blockchain framework developed for enterprise use cases. It features a modular architecture that allows for the customization of components such as consensus mechanisms, identity services, and data storage. HF is particularly well-suited for scenarios requiring strong and fine-grained access controls, privacy, and scalability.

\subsection{Substrate}

Substrate \cite{wood_polkadot_nodate}, developed by Parity Technologies, is a highly flexible blockchain development framework designed to create customized, application-specific blockchains. While it is the foundation of the Polkadot network, Substrate can also function independently, offering unparalleled adaptability for developers aiming to build blockchains with tailored functionalities and consensus mechanisms.

\subsection{Blockchain Interoperability}

Researchers have approached interoperability from multiple perspectives, each highlighting a distinct aspect of this multifaceted challenge.
Several studies conceptualize interoperability as the capability to share data and execute transactions seamlessly across distinct blockchain systems, enabling functionalities such as cross-chain transfers and unified application operations \cite{khan2021towards, belchior_survey_2022}. Other works emphasize technical processes, such as the secure exchange, verification, and validation of information across heterogeneous platforms \cite{hardjono_toward_2020}, or focus on achieving semantic consistency, ensuring data remains valid and interpretable across interconnected systems \cite{koens_assessing_2019}. These viewpoints collectively underscore the technical and operational diversity inherent in blockchain interoperability.

For permissioned blockchains, interoperability introduces additional complexities, especially concerning governance and privacy. Researchers have defined it as the ability to execute and validate transactions across separate networks while preserving data confidentiality and adhering to governance protocols \cite{abebe_enabling_2019}. Additionally, frameworks seek to develop standardized protocols to connect blockchains with differing consensus mechanisms, data structures, and smart contract frameworks, enabling coordinated functionality across heterogeneous systems \cite{siris_interledger_2019}. Achieving effective interoperability is critical for unlocking the broader potential of blockchain technology.

\section {Related Works}

The increasing demand for interoperability among blockchain networks has led to the development of various solutions enabling communication and asset transfers across heterogeneous systems. These frameworks differ in their methodologies, technologies, and target network types, ranging from permissioned to permissionless blockchains.

\subsection{Interoperability Frameworks for Permissionless Blockchains}

\subsubsection{Polkadot}

Polkadot is a leading interoperability framework that enables secure communication and value transfers between heterogeneous blockchain networks through its \textit{relay chain} architecture \cite{wood_polkadot_nodate}. The relay chain acts as a central hub connecting customizable \textit{parachains}, each tailored for specific applications. This shared security model ensures consistency and reliability across connected chains. However, Polkadot primarily focuses on permissionless networks, and its parachains can independently implement their own logic, allowing participants to inject malicious code into the system or cause collusion.

\subsubsection{Cosmos}

This platform \cite{cosmos2019} facilitates interoperability through its Inter-Blockchain Communication (IBC) protocol, connecting independent blockchains, called \textit{zones}, to the central \textit{Cosmos Hub}. This architecture preserves the sovereignty of zones while enabling cross-chain token transfers and message exchanges. Although Cosmos supports both permissioned and permissionless systems, its primary focus lies on permissionless networks, particularly in decentralized finance (DeFi). Challenges arise when integrating highly customized permissioned networks due to differences in governance structures and compliance requirements, limiting its utility in enterprise environments.

\subsection{Interoperability Between Permissioned Blockchains}

\subsubsection{Hyperledger Cacti}

Hyperledger Cacti \cite{hyperledgercacti2025} is specifically designed to connect permissioned blockchains, such as HF, Corda, and Quorum. It adopts a modular architecture with blockchain connectors that facilitate cross-chain communication without altering the underlying network protocols. This enterprise-focused solution addresses governance, security, and compliance needs in private blockchain ecosystems. Despite its effectiveness for permissioned networks, Cacti does not support interoperability with
permissionless systems.

\subsection{Solutions Bridging Permissioned and Permissionless Networks}

\subsubsection{Axelar}

Axelar \cite{axelar2021} supports secure cross-chain communication through a decentralized validator network. Its Universal Cross-Chain Communication (UCC) protocol facilitates data and asset transfers across blockchains with minimal modifications to underlying networks. Axelar is well-suited for decentralized finance (DeFi) platforms but introduces additional trust assumptions through its reliance on validators, which may pose challenges in highly secure or regulated environments.

\subsubsection{Chainlink’s Cross-Chain Interoperability Protocol (CCIP)}

Chainlink’s CCIP \cite{breidenbach_chainlink_nodate} employs decentralized oracles to facilitate secure data exchange across heterogeneous blockchains. While promising, the protocol’s reliance on oracle networks introduces trust dependencies that may conflict with the privacy and compliance needs of permissioned blockchains.

\subsection{Privacy-preserving techniques for cross-chain platforms}
Several studies have addressed the privacy and security of user data. For instance, in \cite{zamyatin2019xclaim}, they proposed Xclaim, a generic framework for exchanging assets between two blockchains using cryptocurrency-backed assets. In \cite{han2025p2c2t}, they propose P2C2T, which is a cross-chain transfer scheme focused on providing atomicity, unlinkability, indistinguishability, noncollateralization, and minimization of required functionalities. Although their solution addresses common security and privacy issues in the area of cross-chain asset transfer, they do not particularly discuss attacks emerged from the nature of blockchain networks such as collusion or deanonymization attacks. Moreover, the existing solutions primarily focus on the atomic swapping, which is a technique to transfer assets between two homogeneous blockchains. As a result, they do not cover network-level security concerns that arise when connecting a permissioned blockchain to a permissionless blockchain.

\subsection{Challenges and Gaps in Existing Solutions}

In contrast to prior work, which focuses either on permissioned-to-permissioned or relay-chain solutions, our contribution is the first to integrate privacy-preserving relay obfuscation with permissioned-to-permissionless communication. This combination of hybrid encryption, source-obfuscation, and security analysis for this specific setting is novel. Privacy in distributed environments can be preserved through several mechanisms, each characterized by distinct advantages and limitations. The four predominant approaches include digital signatures, zero-knowledge proofs, trusted execution environments, and homomorphic encryption \cite{yin2023survey}. Among these, digital signatures represent the most straightforward technique to implement and do not impose additional trust assumptions on the system. Consequently, this work adopts digital signatures as the privacy-preserving mechanism within the proposed architecture.

\section{Architecture}

\begin{figure*}[ht]
    \centering
    \includegraphics[width=\textwidth]{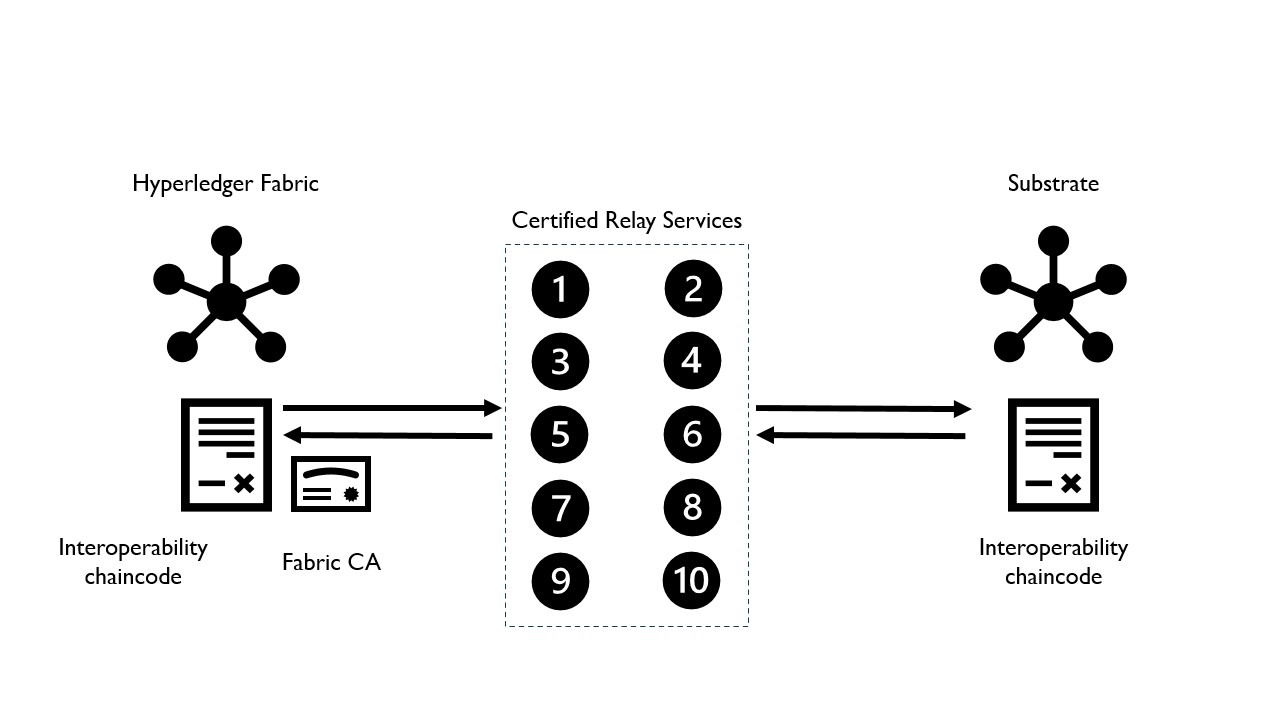}
    \caption{System Architecture Overview}
    \label{fig:architecture}
\end{figure*}

This section describes the key features and components of the proposed architecture that facilitate interoperability between a permissioned blockchain (HF) and a permissionless blockchain (Substrate).

\subsection{System Components}

The architecture introduces the following critical components to establish interoperability, as illustrated in Figure \ref{fig:architecture}:

\begin{itemize}
    \item \textbf{Relay Service}: Acts as an intermediary between the two blockchains. The relay service is blockchain-agnostic and
    capable of connecting any Substrate-based blockchain to
    an HF-based network.
    \item \textbf{Cross-Chain Smart Contracts}: Enable network connectivity by storing configuration information, managing relay installation, and governing transaction flow. These contracts eliminate the need for external entities in network discovery, reducing complexity and enhancing maintainability.
    \item \textbf{Encryption and Signing Keys}: Used to encrypt and sign transaction requests. Each relay node must possess its own keys to participate in the network.
    \item \textbf{Certificate Authority (CA)}): Issues certificates for relay nodes and manages their membership within the system.
\end{itemize}

\subsection{Transaction Flow}

\begin{figure*}[ht]
    \centering
    \includegraphics[width=\textwidth]{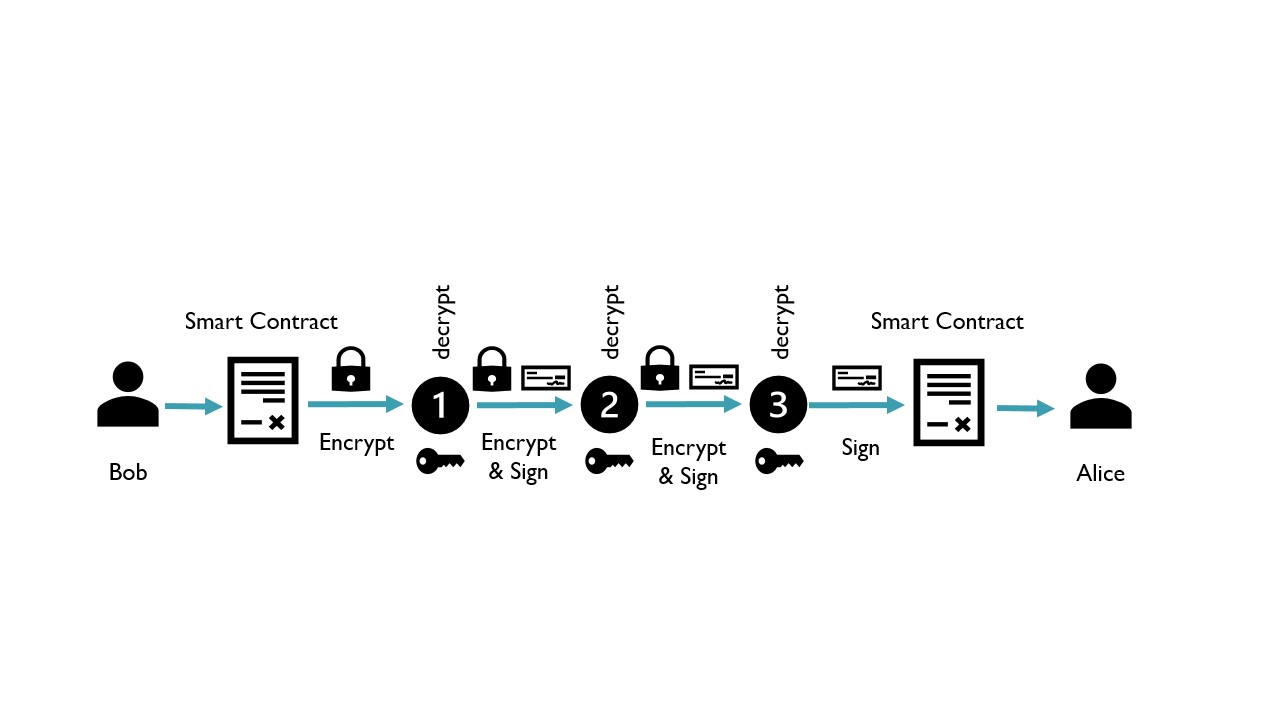}
    \caption{System Transaction Flow}
    \label{fig:transaction_flow}
\end{figure*}

The transaction flow of our architecture is shown in Figure \ref{fig:transaction_flow}. Alice and Bob are users of HF and Substrate, respectively. When Bob initiates a transaction targeting Alice, the request is first received by HF’s smart contract. The Cross-chain smart contract maintains a list of available nodes along with their public keys. It randomly selects a node based on availability and forwards the message, encrypting it with the destination node’s public key.

Upon receiving the message, the selected node decrypts the transaction metadata and subsequently re-encrypts it using the next node’s public key. To ensure data authenticity, the message is signed with the current node’s private signature key prior to forwarding. The next node verifies the signature, decrypts the message, and re-encrypts it for the subsequent node.

The key point here is the anonymity of the user’s identity
and message. Unlike the transaction data, the user identity is
encrypted using the destination smart contract’s public key,
and therefore only to be decrypted by it. As a result, even if the intermediary nodes attempt to act maliciously, they cannot
breach the user’s identity. When sending a request from Bob to Alice, the user’s identity is encrypted using the destination smart contract’s public key, and then decrypted once it reaches the destination. Throughout the forwarding process, none of the relay nodes can read the transaction data or the user who initiated the
transaction.

If Alice, as a Substrate user, wanted to initiate a transaction
against Bob, who is an HF user, the relay nodes would go through the same steps to process the request.

The requirement for protecting transaction data arises from the fact that the relay service functions as an external entity responsible for bridging a permissioned blockchain with a permissionless blockchain. According to the privacy policies of permissioned blockchains, no unauthorized entity should be able to observe user behavior or access sensitive data.

Although relay services are certified, they are not owned by HF and therefore should not be granted access to any information beyond what is strictly necessary to facilitate interoperability between the blockchains.

\subsection{Blind Signature}
We used blind signatures \cite{chaum1983blind} to prevent the signers (smart
contracts and relay nodes) from seeing the content of the
message. Let:

\begin{itemize}
    \item $m$: the message,
    \item $b$: a blinding factor,
    \item $E$: a public key,
    \item $D$: a private key.
\end{itemize}

The user blinds the message:
\[
m' = b \cdot m \pmod{n}
\]

The signer computes the signature:
\[
s' = D(m') = (m')^d \pmod{n}
\]

The user unblinds it:
\[
s = s' \cdot b^{-1} \pmod{n}
\]

Now, $s$ is a valid signature on $m$:
\[
E(s) = m
\]

This safeguards the user's intent and identity while simultaneously authorizing the action.

\subsection{Ring Signature for Relay Nodes}

Each relay node signs its relayed message using a ring
signature scheme, allowing a verifier to confirm the message
was signed by someone in a group $R = \{P_1, P_2, \ldots, P_n\}$,
without revealing who.

Let:
\begin{itemize}
    \item $M$: the message,
    \item $\sigma$: the ring signature,
    \item $\text{Verify}_R(M, \sigma)$: the ring signature verification algorithm.
\end{itemize}

The ring signature satisfies:
\[
\text{Verify}_R(M, \sigma) = \text{true}
\]
yet does not reveal the identity of the signer in $R$.

\subsection{Relay Node Mixing (Onion Routing)}

Relay nodes form a mixing layer to conceal the origin of messages.  
Each message is forwarded randomly through multiple nodes,
with each node adding a layer of signature using ring signatures.
We employed this technique to ensure that only the last node is visible to
the current node.

Each hop adds delay and anonymization:
\[
\text{User} \;\;\to\;\; \text{Relay}_i : \text{Enc}_{R_i}(M)
\]
\[
\text{Relay}_i \;\;\to\;\; \text{Relay}_{i+1} : \text{RingSign}_R(M)
\]

\subsection*{Cross-Chain Token Transfer Flow}

\noindent\textbf{Step-by-step Process}
\begin{enumerate}
    \item User constructs a blinded token transfer message $m'$.
    \item The message is routed through a relay network using
    ring signatures.
    \item Fabric signer receives $m'$ and returns a blind signature
    $s'$.
    \item User unblinds to obtain $s$ on $m$.
    \item User sends $(m, s)$ to Substrate (again via relays).
    \item Substrate smart contract verifies:
    \[
    E(s) = m
    \]
    \item Token is credited to the Substrate account.
\end{enumerate}

\subsection{Source Anonymization}

Another challenge addressed by the proposed architecture is the obfuscation of transaction origins to mitigate \textit{de-anonymization attacks}. Such attacks, as described in \cite{zhang_txspector_nodate}, exploit the susceptibility of the initial broadcasting node, allowing adversaries to infer transaction details. If a relay node is compromised, confidential transaction data may be exposed, thereby undermining both privacy and security.

To counteract this threat, the architecture incorporates \textit{source-obfuscation protocols}, specifically \textit{Clover} and \textit{Dandelion++}. While these protocols were originally designed for node-to-node communication within a single blockchain network, they have been adapted here to operate within the relay service, facilitating cross-chain interoperability. The system evaluates their effectiveness in preserving privacy while maintaining efficiency.

These protocols forward messages across multiple relay nodes, ensuring that at each step, a node is only aware of its immediate predecessor and successor. This design enhances privacy, as an adversary compromising a single relay node remains unable to determine either the transaction's origin or final destination.

From a security perspective, \textit{compromising the entire transaction flow requires an adversary to infiltrate all participating relay nodes}. Let \( f \) denote the number of forwarding steps (equivalent to the number of relay nodes involved in message propagation) and \( n \) denote the total number of available relay nodes. The probability of an adversary successfully compromising any single participating node is therefore \( f/n \).

Given that each relay node has visibility only over its adjacent nodes, the compromise of a single node reveals information about at most three nodes within the forwarding process: \textit{the preceding node, the current node, and the succeeding node}. In the worst-case scenario, an adversary capable of breaching every alternate node in the forwarding chain could reconstruct the entire transaction path. This implies that to achieve full path reconstruction, the adversary must compromise at least \( f/3 \) nodes. Consequently, the probability of an adversary successfully breaching all participating nodes in a given transaction is given by:

\begin{equation}
    p(B_{Tx}) = \prod_{i=0, j=0}^{\frac{3f}{n}} \frac{f - j}{n - i}
    \label{formula_1}
\end{equation}

Where \(p(B_{Tx})\) denotes the probability of breaching transaction $Tx$ and \( \frac{f - i}{n - i} \) represents the probability of breaching the \( i \)-th node. The value of \( j \) depends on the number of nodes the adversary has compromised. We can calculate \( j \) using the following formula:\\

\begin{equation}
    j =
    \begin{cases} 
    j, & \text{if } n_i \notin N_{FP} \\
    j + 1, & \text{if } n_i \in N_{FP}
    \end{cases}
\end{equation}

where $N_{FP}$ denotes the set of all nodes that were part of the forwarding path.

Every time a node is breached, the pool of untampered nodes shrinks by one. As we stated earlier, each breached node will expose the adjacent nodes in the forwarding path, causing three nodes to be breached at the end (the previous, the current, and the next node). This means that an adversary must breach at least $\frac{n}{3}$ of the nodes. In conclusion, breaching $\frac{f}{\frac{n}{3}}$ or $\frac{3f}{n}$ of the nodes can cause transaction information leakage.

\paragraph{Key Takeaways of the Security Analysis.}
Based on the formula \ref{formula_1}, we can conclude that:
\begin{itemize}
    \item Initially, $j = 0$, so the fraction starts at $\frac{f}{n}$.
    \item As more nodes in $N_{FP}$ are breached, $j$ increases, reducing $f - j$, which makes the probability smaller.
    \item Simultaenously, the denominator $n - i$ shrinks, making the fraction more sensitive as $i$ grows.
\end{itemize}

As a result, we can entail that
\begin{itemize}
    \item The higher the value of $f$,the higher the risk of data breach.
    \item A network with more nodes is less likely to breach data.
    \item If a breach occurs early (low $i$), it can spread quickly due to the $3f$ exposure effect.
\end{itemize}
Based on this analysis, the higher value of $\frac{f}{n}$ poses a bigger threat to the system's privacy. However, the anonymity of the system is preserved since the adversary cannot infer the transaction origin to associate the data with a specific user. As a result, there is a trade-off between the anonymity of the transactions and the data breach.

\subsection{Source-Obfuscation Mechanisms}
We incorporated Dandelion++ and Clover protocols separately in our system to provide transaction anonymity and further enhance user trust.

 \begin{figure}[ht!]
    \centering
    \includegraphics[width=\columnwidth]{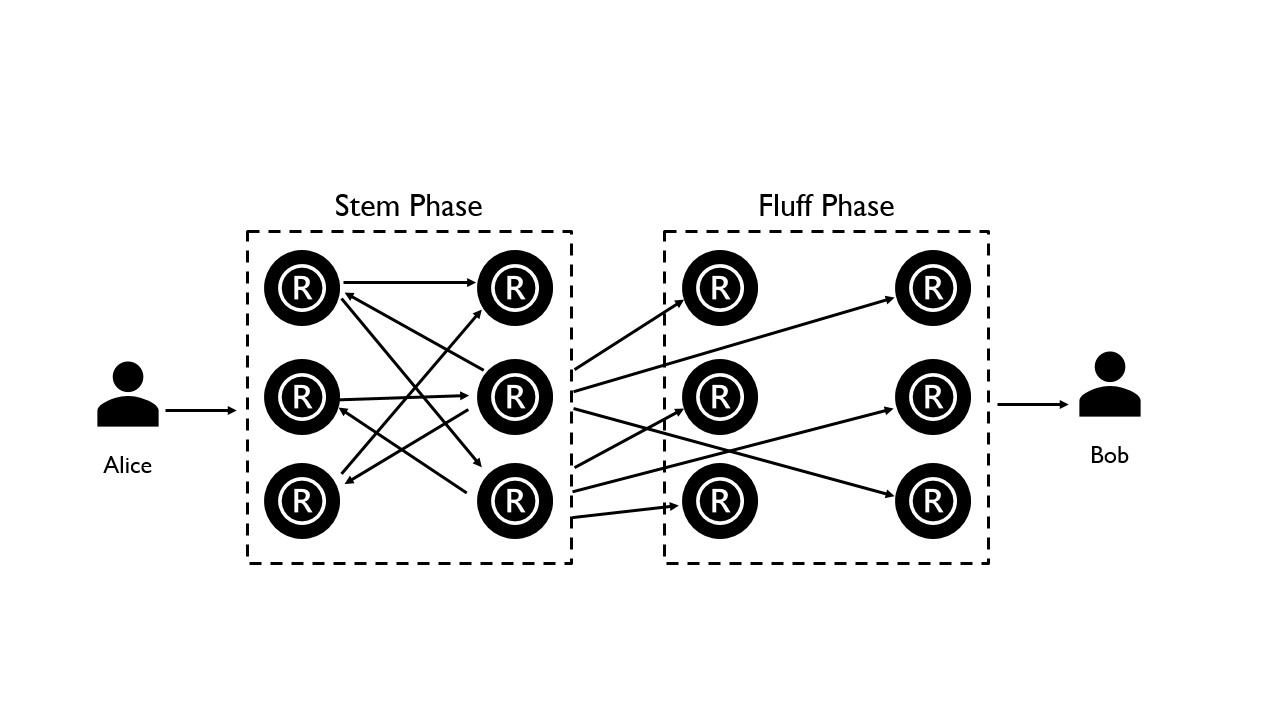}  
    \caption{Dandelion++ Transaction Flow}
    \label{fig:dandelion}
\end{figure}

\subsubsection{Dandelion++}

Dandelion is a routing protocol designed to obfuscate which node originated a transaction. The protocol has two phases: a \textit{stem} phase where a transaction is forwarded along a random path of nodes (ideally with no loops/backtracking), followed by a \textit{fluff} (broadcast) phase where it is gossiped to the wider network. It is to ensure that an outside observer seeing the broadcast cannot easily link it back to the true origin, because the stem path conceals the
source among many possible candidates.

\subsubsection{Clover}

\begin{figure}[ht!]
    \centering
    \includegraphics[width=\columnwidth]{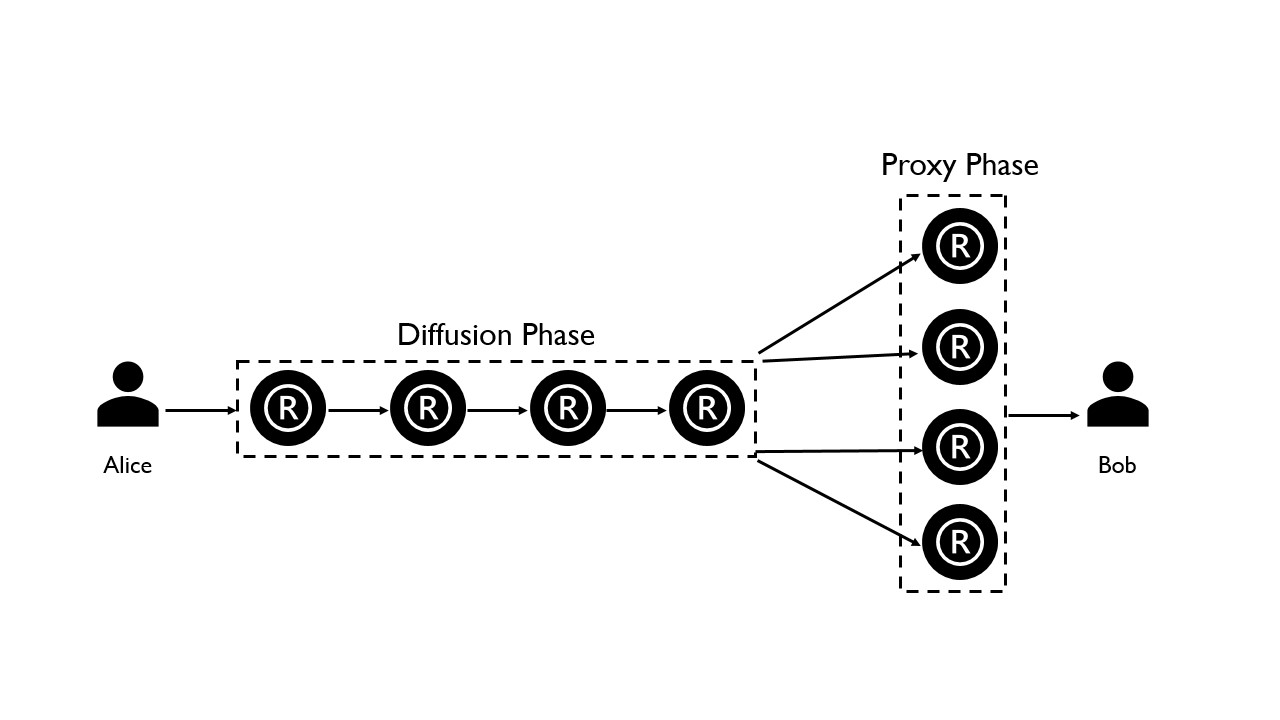}  
    \caption{Clover Transaction Flow}
    \label{fig:Clover}
\end{figure}

Clover is a transaction relay protocol designed to obscure the origin of a broadcast in a P2P network. Like Dandelion++, it seeks to break the link between a sender’s IP address and the point of broadcast. The protocol operates in two phases: a diffusion phase, where a transaction is sent to multiple randomly selected relay nodes rather than a single path, enhancing resilience against path inference attacks; and a proxy phase, where the transaction continues to hop between nodes until a probabilistic condition is met, at which point it is submitted to the target blockchain’s smart contract. By distributing the transaction through several intermediaries and avoiding a predictable forwarding path, Clover reduces the risk of adversaries identifying the true origin node.

\subsection{Relay Service and Network Integration}
The relay nodes are configured to interact with both blockchain networks securely.

\begin{itemize}
    \item \textbf{HF Integration}: The relay service connects to HF through a dedicated interoperability channel. Channels in HF provide a private layer of communication, isolating sensitive transactions and ledger data from the rest of the network. This adds an extra layer of confidentiality and privacy to the entire system.
    \item \textbf{Certificate-Based Authentication}: Any relay attempting to join the network must possess a valid certificate issued by HF's Certificate Authority (CA).
\end{itemize}

\section{Security and Privacy Analysis}

To evaluate the privacy and security of our system, we
adopt the STRIDE threat model \cite{shostack2014threat}, which identifies six
primary categories of security threats in software systems:
\textit{spoofing}, \textit{tampering}, \textit{repudiation}, \textit{information disclosure}, 
\textit{denial of service}, and \textit{elevation of privilege}. 
Aside from these threats, we discuss \textit{collusion}, which is defined as any malicious
coordinated behavior of a group of users aimed at gaining
undeserved benefits \cite{ciccarelli2011collusion}. Collusion attacks are common security concerns within decentralized environments.

\subsection*{Spoofing}
According to the STRIDE framework, spoofing threats are primarily concerned with the compromise of authentication mechanisms. These threats involve an attacker attempting to bypass identity verification protocols in order to impersonate a legitimate user or process \cite{rowe2011role}.

In the context of our system, spoofing may occur at the
level of relay nodes. An attacker could attempt to impersonate
a legitimate relay node in order to intercept messages, thereby
compromising both data integrity and confidentiality. To mitigate this threat, each relay node is required to present a valid
certificate before being permitted to participate in the network.
Nodes that fail to provide a valid certificate are denied access
and are thus unable to intercept or relay messages within the
system.

\subsection*{Tampering}
Tampering refers to the unauthorized modification of data 
either in transit or at rest, with the intent to manipulate, disrupt, or corrupt system behavior \cite{shostack2014threat}. 

To mitigate this threat, all relay nodes and blockchain interfaces
are designed to verify message signatures prior to processing. Any message that is unsigned or bears an invalid signature is rejected and logged for the purpose of detecting potential malicious activity. To ensure both confidentiality and integrity, we employed an \textit{authenticated encryption with associated data (AEAD)} scheme such as AES-GCM. This scheme ensures that decryption fails automatically if tampering is detected.

Additionally, relay node activities are recorded using an append-only, hash-chained logging mechanism, which prevents unauthorized modifications to the logs and supports post-incident auditing.

\subsection*{Repudiation}
Repudiation is a threat wherein a party denies involvement in a transaction, message, or event, particularly when there is no reliable evidence (e.g., logs or digital signatures) to prove otherwise. 

In our system, each message is digitally signed by its sender, providing cryptographic proof of origin and ensuring non-repudiation. This mechanism is applied uniformly to both users and relay nodes, thereby preventing any entity from denying the initiation or transmission of a message. To further mitigate repudiation risks, all transaction-related information is persistently logged, which makes it infeasible for an adversary to erase or alter the transaction history.

\subsection*{Information Disclosure}
Information disclosure refers to \textit{the unauthorized exposure of information to a component for which it is not intended} \cite{rouland2021specification}. To mitigate this threat, all messages are encrypted using the recipient’s public key, ensuring that only the intended receiver can decrypt and access the message content.

\subsection*{Denial of Service}
According to Tandon et al. \cite{tandon2020survey}, \textit{a Denial of Service (DoS) attack aims to disrupt the availability of a targeted system by overwhelming it with a flood of
illegitimate requests, rendering it inaccessible to legitimate users}. In our system, a malicious user or relay node could potentially overwhelm a target node, thereby rendering it unresponsive.

To address this threat, we implement rate-limiting mechanisms that restrict the number of requests a node can receive within a defined time window. If the request threshold is exceeded, additional requests are discarded, and an acknowledgment is sent to inform the sender of the failure. Although the rate-limiting mechanism is configured, it is currently deactivated to allow performance evaluation without prematurely rejecting transactions.

\subsection*{Elevation of Privilege}
Elevation of privilege (EoP) occurs when \textit{an attacker gains unauthorized access to higher-level permissions, allowing them to perform actions beyond their intended rights} \cite{kujanpaaAutomating}. To mitigate this threat, we have implemented a monitoring mechanism designed to detect anomalous or unauthorized behavior from system participants. Additionally, all user requests and node inputs are validated through smart contracts, which serve as a safeguard against injection attacks and unauthorized privilege escalation.

\section{Evaluation}

The evaluation of the proposed system focuses on the performance analysis of the forwarding approaches as well as a simulated analysis of the collusion attacks with respect to deanonymization probability and anonymity set size. The forwarding approaches are compared with another forwarding strategy known as the Shortest Ping approach. This
method selects the relay node with the shortest response time, prioritizing performance over privacy and security considerations.

\subsection{Experimental Setup}

The experiments \footnote{\href{https://github.com/anonymous-researcher-77/ARES_2025}{GitHub}} were conducted using the following blockchain environments:
\begin{enumerate}
    \item \textbf{Substrate-Based Network}: Configured using the Substrate Contracts Node, which incorporates pallets and essential interfaces for deploying and interacting with smart contracts. This setup ensures that the system operates with the base Substrate node independently of additional pallets
    \item \textbf{Hyperledger Fabric (HF) Test Network}: Used as the permissioned blockchain environment, configured with channels and a Certificate Authority (CA) to facilitate secure communication.
    \item \textbf{Relay Service}: Consists of 100 relay nodes that interact with both blockchain and with one another to interconnect HF with Substrate.
\end{enumerate}

\subsection{Forwarding Time}

\begin{figure}[t]
    \centering
    \includegraphics[width=0.8\linewidth]{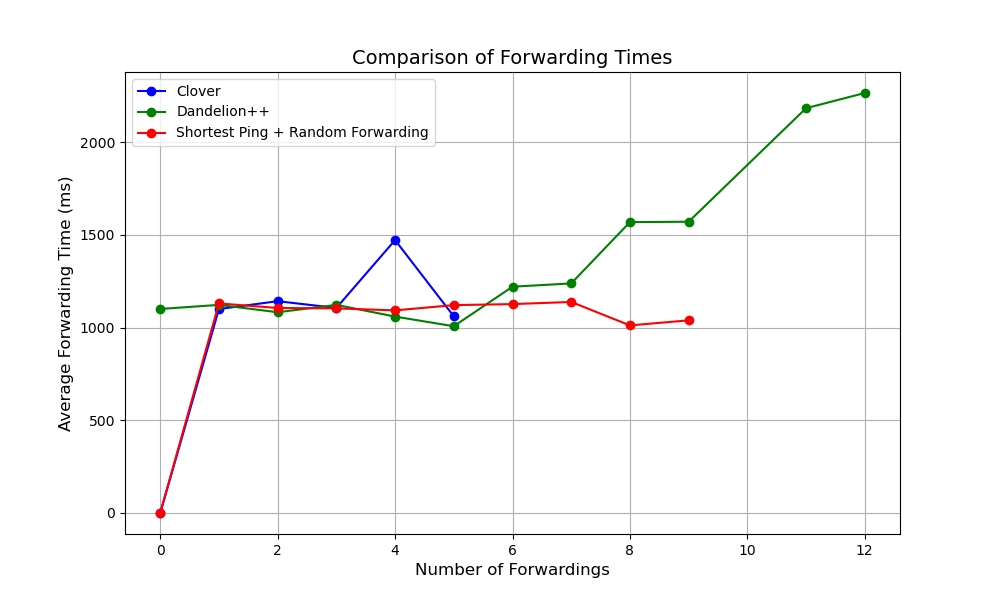} 
    \caption{Comparison of average forwarding times with respect to the number of forwardings}
    \label{fig:forwarding_times}
\end{figure}

The results underscore the trade-offs between performance and privacy across the forwarding approaches. Clover strikes a balance between forwarding time and the number of forwardings, making it well-suited for applications that require both efficiency and moderate privacy. Dandelion++ prioritizes robust source obfuscation and privacy, albeit at the cost of higher forwarding times and increased relay involvement, making it ideal for scenarios where anonymity is paramount.

\subsection{Throughput}

\begin{figure}[t]
    \centering
    \includegraphics[width=0.8\linewidth]{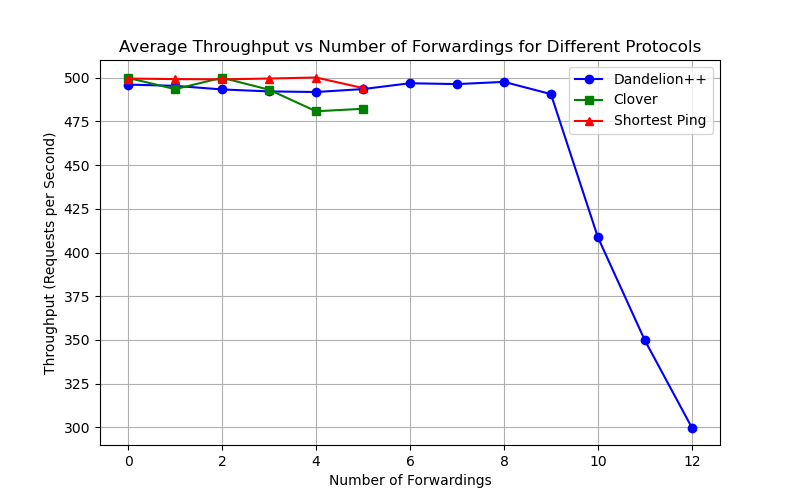} 
    \caption{Comparison of the average throughput with respect to the number of forwardings}
    \label{fig:throughput}
\end{figure}

We evaluated the throughput of our system for each forwarding approach independently. Before presenting the results, we first define throughput in the context of our system.

The throughput of a relay node is defined as the number of transactions a node can process per second. The overall throughput of the relay service, comprising \textit{n} relay nodes, is the total number of transactions processed collectively by all nodes. Theoretically, the system throughput is expressed as follows:

\begin{equation}
    TP_R = \sum_{i=1}^{n-1} TP_{r_i}
    \label{formula_3}
\end{equation}

where \( TP_R \) represents the total system throughput, and \( TP_{r_i} \) denotes the throughput of the \( i \)-th relay node.

As shown in Figure \ref{fig:throughput}, Clover demonstrates a performance comparable to the \textit{Shortest Ping} approach, whereas the throughput of Dandelion++ declines significantly when the number of forwarding steps exceeds ten.

\subsection{Availability}
\begin{figure}[t]
    \centering
    \includegraphics[width=0.8\linewidth]{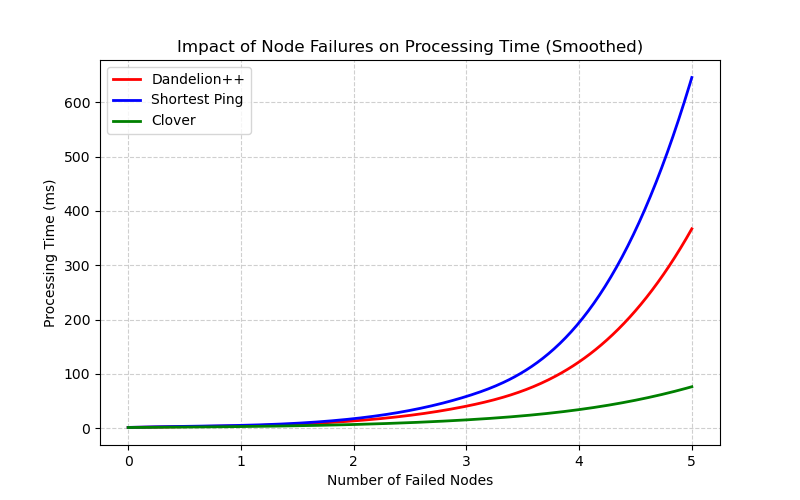} 
    \caption{Comparison of average processing times with respect to the number of unavailable nodes}
    \label{fig:node_failure_delay}
\end{figure}

The system's availability is evaluated by deliberately deactivating a subset of relay nodes to analyze their impact on network performance. Figure \ref{fig:node_failure_delay} illustrates the processing time of transactions as a function of the number of unavailable nodes.

The processing times are computed by averaging the transaction processing time across 6000 requests for each failure scenario. The graph compares the performance of each forwarding approach to determine which method demonstrates greater resilience to node failures.

The evaluation results indicate that all approaches exhibit similar performance when only a small number of nodes fail (fewer than three). However, as the number of failed nodes increases, the processing time escalates exponentially for both approaches. Between Dandelion++ and Clover, Clover demonstrates a slower increase in processing time. This behavior can be attributed to Clover's ability to broadcast requests to a sufficient number of nodes, thereby enhancing performance while maintaining anonymity.

\begin{figure}[t]
    \centering
    \includegraphics[width=0.8\linewidth]{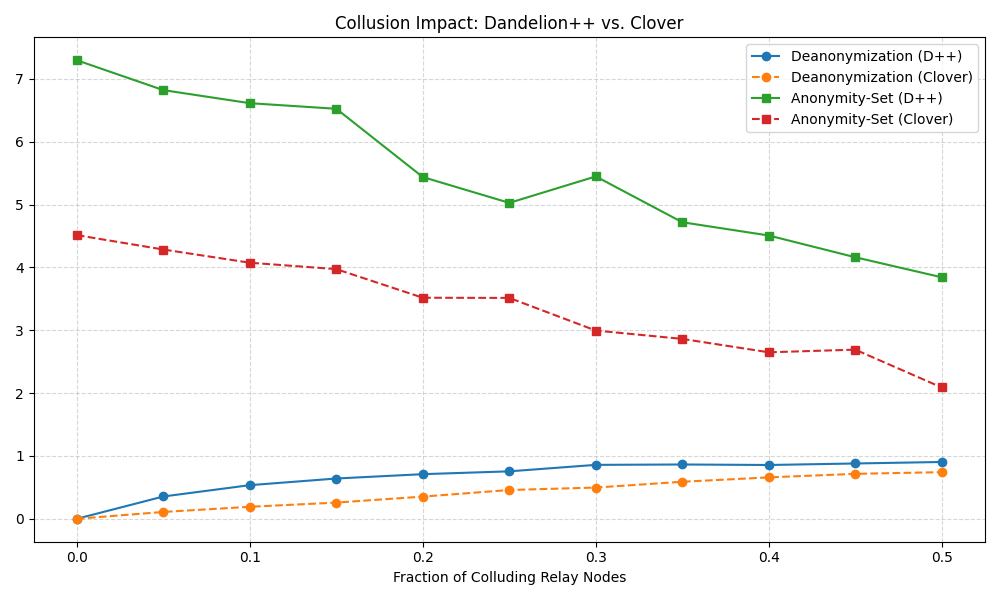} 
    \caption{Comparison of Dandelion++ and Clover under varying collusion ratios}
    \label{fig:collusion}
\end{figure}

\subsection{Impact of Collusion on Anonymity and Deanonymization}

Figure \ref{fig:collusion} presents a comparative analysis of the resilience of Dandelion++ and Clover against collusion attacks, quantified in terms of (i) probability of deanonymization, and (ii) average anonymity set size, as the fraction of colluding relay nodes
increases from 0 to 0.5.

\subsection*{Deanonymization Probability}
Dandelion++ exhibits a steeper increase in deanonymization probability, starting near
zero and approaching $90\%$ as half the network colludes. In contrast, Clover remains substantially more resistant, with the deanonymization rate staying below $70\%$ even under $50\%$ collusion. This indicates that Clover’s probabilistic random
forwarding yields more robust obfuscation than Dandelion++’s deterministic stem-fluff mechanism in adversarial settings.

\subsection*{Anonymity Set Size}
Dandelion++ maintains a higher anonymity set size overall, starting around $7$ and dropping gradually as collusion increases. However, Clover exhibits a
more stable profile, with smaller anonymity sets (starting at $4.5$) but a slower decline. This suggests Dandelion++ benefits from broader candidate origin sets in benign conditions, while Clover sacrifices initial size for better resilience against
targeted inference.

\subsection*{Takeaway}
Clover achieves a better trade-off between deanonymization resistance and anonymity-set stability in the presence of adversarial nodes, making it more suitable for deployment in partially compromised networks. Dandelion++, while initially offering larger anonymity sets, becomes increasingly fragile as collusion intensifies.

\section{Conclusion}

This work presented a blockchain interoperability framework built upon a hybrid Substrate-Fabric architecture, enabling secure and anonymous cross-chain communication via relay nodes. By incorporating cryptographic primitives and privacy-preserving strategies, the system ensures end-to-end encryption, transaction unlinkability, and protection against
passive and active adversaries. To assess the anonymity guarantees of the framework, we integrated and simulated two prominent source-obfuscation
protocols: Dandelion++ and Clover. Our evaluation demonstrated that while both protocols enhance transaction-level privacy under adversarial models such as collusion, Clover consistently offered better resilience and lower deanonymization probabilities. Additionally, performance metrics—such as throughput, average forwarding delay, and availability—further highlighted Clover’s advantage in dynamic network environments.

\end{document}